\documentclass[flushrt,preprint]{aastex}
\tighten
\eqsecnum

\input psfig

\begin{document}

\title{The Expected Properties of Dark Lenses}
\author{David Rusin} 
\affil{Harvard-Smithsonian Center for Astrophysics, 60 Garden St., MS-51,
Cambridge, MA 02138}

\begin{abstract}

We investigate the properties of multiple-image gravitational lens
systems formed by dark matter halos. Both the Navarro-Frenk-White
(NFW) and Moore et al.\ mass profiles are considered, and the effects
of quadrupole perturbations to the lensing potential are taken into
account. The systems produced by dark halos exhibit two generic
properties that serve as powerful benchmarks for scrutinizing
individual lens candidates: small flux ratios between the two
brightest images in triples, and detectable odd images. In contrast,
most currently ambiguous quasar pairs consist of two components with
flux ratios $\geq$ 3:1. Such systems are statistical outliers in the
expected distribution of dark lenses, and are therefore likely to be
binary quasars.

\end{abstract}


\section{Introduction} 

Observational evidence indicates that virtually all galaxies are
embedded in dark matter halos, but do all halos host galaxies?
Differences in the slope of the mass function for galaxies (e.g.,
Blanton et al.\ 2001) and dark halos (Press \& Schechter 1974) suggest
that a substantial fraction of halos may be empty, particularly at
lower masses (e.g., Nagamine et al.\ 2001). Because cold dark matter
(CDM) simulations predict cuspy halos (e.g., Navarro, Frenk \& White
1997; Moore et al.\ 1999b; Jing \& Suto 2000) that are sufficiently
concentrated to multiply image background sources (Li \& Ostriker
2002; Wyithe, Turner \& Spergel 2001), strong gravitational lensing
may be a powerful tool for detecting empty dark matter halos.

The identification of gravitational lens candidates is based on the
presence of two or more closely-separated quasars at the same
redshift. In cases where a galaxy, group or cluster is responsible for
the multiple imaging, the lens hypothesis can be unambiguously
confirmed by directly detecting the lensing mass through optical or
infrared observations. This is clearly not possible for dark
lenses. A thorough investigation of the quasar components is therefore
necessary to determine whether a system is a genuine dark lens, or a
pair of physically distinct quasars. The lensing hypothesis can be
robustly rejected by finding some spectral or photometric property
that the quasars do not share in common. For example, a pair with only
one radio-bright component (an $O^2R$ system, see Kochanek, Falco \&
Mu\~noz 1999) is inconsistent with gravitational lensing, and
must therefore be a binary quasar. The inability to uncover
differences in the quasars is not compelling evidence for lensing,
however, as similarities can be at least partially attributed to the
common formation history and cosmic environment of binaries. Unless
the presence of three or more quasar components makes the lensing
identification obvious, a measurement of the time delay or a detection
of correlated milliarcsecond-scale radio substructure would likely be
necessary to prove that a system is a dark lens.

A number of wide-separation ($\Delta \theta \geq 3''$) quasar pairs
with no detectable lensing mass have been reported (e.g., Hawkins
1997; Mortlock, Webster \& Francis 1999), and their true nature has
been the subject of continuing debate. Several of the quasar pairs are
$O^2R$ systems (e.g., MGC 2214+3550; Mu\~noz et al.\ 1998), and
therefore binaries. For others the lensing hypothesis has been
rejected based on incompatible spectral line widths (e.g., CTQ 839;
Morgan et al.\ 2000) or X-ray properties (e.g., Q2345+007; Green et
al.\ 2002). Yet, there exist $\sim 10$ systems for which the
identification remains ambiguous. Statistical arguments suggest that
most of these pairs are likely to be binary quasars (Kochanek et al.\
1999). The scarcity of radio-bright/radio-bright dark lens candidates
is particularly difficult to reconcile with a sample dominated by
lenses, considering how many radio lenses have been discovered. Still,
the existence of several dark lenses is allowed by statistical
constraints, and such arguments cannot determine the nature of any one
source. It is therefore desirable to find additional criteria by which
to scrutinize {\em individual} dark lens candidates.

Most studies of strong gravitational lensing by dark matter halos have
focussed on their contribution to the optical depth (Keeton \& Madau
2001; Li \& Ostriker 2002; Sarbu, Rusin \& Ma 2001), rather than the
properties of the lens systems produced. This paper bridges the gap,
and will serve as a 'field guide' for dark lenses. In \S2 we
investigate lenses formed by plausible models of dark halos,
quantifying the distribution of magnification ratios between the
two brightest images in triples, as well as the relative flux of the
third image. In \S3 current dark lens candidates are compared to our
model predictions. Section 4 summarizes our findings and discusses
their implications.

\section{Predictions for Dark Lens Models}

Numerical simulations predict CDM halos with cuspy but shallow inner
mass profiles ($\rho \propto r^{-\alpha}$ at small radii, where $1.0 <
\alpha < 1.5$; Navarro et al.\ 1997; Moore et al.\ 1999b). This result
applies to halos in their ``primordial'' state. Cooling baryons will
modify the dark matter profile through adiabatic compression (e.g.,
Blumenthal et al.\ 1986; Keeton 2001; Kochanek \& White 2001),
producing the more centrally-concentrated mass distributions that are
characteristic of lens galaxies (e.g., Cohn et al.\ 2001; Kochanek
1995a; Rusin \& Ma 2001). Inversely, if the baryons have not cooled,
the dark matter halo is expected to retain its primordial profile
(e.g., Navarro, Frenk \& White 1995; Eke, Navarro \& Frenk
1998). Star formation should be greatly inhibited in such halos, so
the mass distributions will be truly ``dark'' (except perhaps at X-ray
wavelengths). The existence of a characteristic mass scale separating
cooled halos with nearly isothermal profiles (efficient lenses) and
non-cooled halos with primordial dark matter profiles (inefficient
lenses; e.g., Wyithe et al.\ 2001) is vital for converting the halo
mass function into the observed distribution of lensed image
separations (Porciani \& Madau 2000; Kochanek \& White
2001). Specifically, halos above this scale must have a much lower
cross section per unit mass to account for the sharp decline in
observed lenses with $\Delta \theta > 3''$. The necessary mass scale
($M_c \simeq 10^{13} M_{\odot}$) agrees well with the predictions of
semi-analytic cooling models (e.g., Cole et al.\ 2000). To produce
arcsecond-scale ($1'' < \Delta \theta < 10''$) lens systems, halos
that retain their shallow primordial profiles require virial masses
$M_{\rm vir} \ga 10^{13.5} M_{\odot}$ (e.g., Li \& Ostriker 2002) --
larger than the cooling scale. Consequently, the formation of
observable multiple-image lenses by uncooled and largely unmodified
CDM halos is self-consistent. Note, however, that the lack of star
formation may not imply an uncompressed halo. In particular, baryons
in halos with $M_{\rm {vir}} < 10^{12} M_{\odot}$ can cool into a disk
that is stable to star formation if the spin parameter is high
(Jimenez et al.\ 1997). Such halos would be both dark and
concentrated. But even if the resulting mass profile were as steep as
isothermal, the limit of $10^{12} M_{\odot}$ corresponds to a lensed
image separation of less than an arcsecond (e.g., Li \& Ostriker
2002). The above arguments therefore suggest that arcsecond-scale dark
lens systems are likely to be produced by shallow CDM mass profiles.

This paper investigates two popular models for cuspy CDM halos: the
Navarro-Frenk-White (NFW; 1997) mass distribution ($\alpha = 1$), and
the steeper ($\alpha =1.5$) profile of Moore et al. (1999b). While the
inner profile slope may be mass dependent (Jing \& Suto 2000), the
above models span the typical range of values found. Spherical mass
models are considered for computational simplicity, but we place them
in constant shear fields (magnitude $\gamma$) to mimic the effects of
deflector ellipticity and environmental perturbations. We focus on
two/three-image configurations, which we term ``triples,'' because
these are the systems most likely to be confused with binary quasars.
The deflector models are analyzed by numerically solving for the image
positions and magnifications ($\mu_i$) of point sources placed on a
uniform grid behind the lens. The distributions of two quantities are
then extracted: the magnification ratio between the two brightest
images ($r_{12} \equiv |\mu_1 / \mu_2|$), and that between the
brightest and third brightest images ($r_{13} \equiv |\mu_1 /
\mu_3|$).

The properties of dark lenses differ significantly from those of the
more concentrated, nearly isothermal ($\rho \propto r^{-2}$) galaxy
lenses. Specifically, the magnification ratio distributions are very
sensitive to the size of the radial critical curve, which increases as
the slope of the inner mass profile is decreased from isothermal
(Blandford \& Kochanek 1987; Wallington \& Narayan 1993; see Kormann,
Schneider \& Bartelmann 1994 for plots). First, source positions
enclosed by only the radial caustic produce triples in which the
second (negative-parity) image resides outside of the radial critical
curve; the central (positive-parity) image resides within it. Large
radial critical curves imply that the outer images tend to form at
similar distances from the lens center, resulting in similar
magnifications and small primary flux ratios $r_{12}$. The third image
can also form farther from the lens center, away from the high central
convergence that would otherwise strongly demagnify it. Consequently,
this image carries more flux, leading to smaller values of
$r_{13}$. Second, the critical structures of shallower mass profiles
are more susceptible to shear (e.g., Wallington \& Narayan 1993),
which allows for the production of naked cusp configurations. Here,
source positions enclosed by only the tangential caustic are lensed
into three images on the same side of the deflector, each with similar
magnification (e.g., Kormann et al.\ 1994). This class of triples is
characterized by very small magnification ratios and bright third
images, thereby reinforcing the above trends.

Flux ratio distributions are affected by magnification bias, which
favors the detection of more magnified lens systems (e.g., Turner
1980). The bias factor depends on the lens model, the number-flux
relation of the sources, and the method of finding lenses. A full
accounting of magnification bias is beyond the scope of this
paper. However, we illustrate the bias effect by simply weighting each
lens system generated on the source plane grid by its total
magnification ($\mu = \sum_i |\mu_i|$). For example, the biased
fraction of lenses with $r_{12} < r_0$ is $[\int \mu(y_1, y_2) R(y_1,
y_2, r_{12}) dy_1 dy_2] / [\int \mu(y_1, y_2) dy_1 dy_2]$, where $R$ =
1 if $r_{12} < r_0$ for a source at $(y_1, y_2)$, and 0 otherwise.
The integral is evaluated over the three-image caustic regions. This
scheme mimics the bias produced by the differential number-flux
relation $N(S) \propto S^{-2}$, which closely approximates the
population of compact radio sources (Rusin \& Tegmark 2001). Faint
optical quasars have a much shallower luminosity function, and hence
they are nearly unbiased (Kochanek 1996). Bright optical quasars are
more strongly biased, as the luminosity function is very steep in that
regime.

\subsection{NFW Profile}

The NFW model is described by the mass density
\begin{equation}
\rho(r) = \frac{\rho_{cr} \bar{\delta}_I} {(r/R_s)  [1 + (r/R_s)]^2}
\end{equation}
where $R_s = R_{200}/c(z)$, $R_{200}$ is the radius within which the
average density is 200 times the critical density $\rho_{cr}$ at
redshift $z$, and $c(z)$ is the concentration parameter. For a flat
$\Omega_{\Lambda} = 0.7$ cosmology, $c_0 \equiv c(0) \simeq 7$ (e.g.,
Bartelmann et al.\ 1998) and evolves as $c(z) = c_0/(1+z)$ (e.g.,
Bullock et al.\ 2001). The density amplitude is $\bar{\delta}_I =
(200/3) c^3 / [\ln(1+c) - c/(1+c)]$. The spherical NFW profile has an
analytic projected surface density and deflection angle (Bartelmann
1996). Its lensing properties depend on the parameter $\kappa_0 =
\rho_{cr} \bar{\delta_I} R_s / \Sigma_{cr}$, where $\Sigma_{cr}$ is
the critical surface mass density. The angular image splitting scale
($\Delta \theta$) is set by the size of the tangential critical
curve. In Fig.~1, $\Delta \theta$ is plotted as a function of
$\kappa_0$ for some typical redshifts and concentrations.

The fractions of NFW triples with a magnification ratio between the
two brightest images of $r_{12} < 1.5$, $2.0$ and $5.0$ are plotted in
Fig.~2. For clarity, we plot the cumulative distributions of $r_{12}$
for $\kappa_0 = 0.10$, $0.15$, and $0.20$ in Fig.~3.  Note that
magnification bias only slightly alters the values, and does not
affect the general trends.  For a spherically-symmetric potential
($\gamma = 0$), almost all triples have $r_{12} \simeq 1$.  Shear
perturbations create a tangential caustic on the source plane,
affecting the distribution of magnification ratios. If the cusp of the
tangential caustic resides within the radial caustic, a single highly
magnified image results for sources near the cusp. Such systems can
have large $r_{12}$, and therefore the fraction of lenses with
$r_{12}$ less than some nominal value is decreased. If the cusp is not
enclosed by the radial caustic, however, the deflector can produce
naked cusp configurations.  Such systems have $r_{12} \la 2$, and
therefore the fraction of lenses with small $r_{12}$ is
increased. Deflectors with smaller $\kappa_0$ are more susceptible to
shear perturbations, as the ratio of the radial to tangential critical
curve increases with decreasing $\kappa_0$ (Bartelmann 1996). Hence,
for a fixed $\gamma$, the fraction of systems with small $r_{12}$
rises sharply below that $\kappa_0$ at which naked cusp configurations
begin to dominate.  This effect is clearly seen in Figs.~2 and 3. The
conclusion is that NFW dark lenses with $1'' < \Delta \theta < 10''$
(using the conversion in Fig.~1) are expected to have flux ratios very
close to unity if either the potentials are nearly spherical, or have
moderate to large shear perturbations ($\gamma \ga 0.10$). But even in
the intermediate regime, systems with high flux ratios are still
rare. For $\gamma = 0.05$, nearly half of arcsecond-scale NFW lenses
will have $r_{12} < 2$, and almost all will have $r_{12} < 5$
(Fig.~2). In comparison, of the lenses produced by a singular
isothermal sphere, 11\% have $r_{12} < 2$ and 44\% have $r_{12} < 5$.

The relative magnification of the third image ($r_{13} \equiv
|\mu_1/\mu_3|$) is examined in Figs.~4 and 5. Plotted in Fig.~4 are
the fractions of NFW triples with $r_{13} < 20$, $50$ and
$100$. Plotted in Fig.~5 are the cumulative distributions of $r_{13}$
for $\kappa_0 = 0.10$, $0.15$, and $0.20$. The third image tends to be
very bright and easily detectable, independent of $\kappa_0$ or
$\gamma$. In almost all cases, at least half of the triples will have
$r_{13} < 20$. The third image is even more detectable in regions of
parameter space dominated by naked cusps, as these systems have three
images of very similar magnification.

\subsection{Moore Profile}

The model of Moore et al. (1999b) is described by the mass density 
\begin{equation}
\rho(r) = \frac{\rho_{cr} \bar{\delta}_{II}} {(r/R_s)^{3/2} [1 +
(r/R_s)^{3/2}]}
\end{equation} 
where $\bar{\delta}_{II} = 100 c^3 / \ln(1+c^{3/2})$, and $c$ is a
factor of a few smaller than the NFW value. There is no analytic form
for the projected surface density or deflection angle. Typical input
parameters, however, produce lensed images at radii that are much
smaller than the arcminute-scale angular break radius derived from
$R_s$. Because lenses will be dominated by the inner cusp, the Moore
profile can be well represented by the power-law mass density $\rho(r)
\propto r^{-1.5}$. The critical structures of power-law models scale
in a self-similar manner with the profile normalization, and hence the
lensing properties are normalization-independent. Trial analyses of
the real Moore profile demonstrate that it is well approximated by the
power law and exhibits the expected normalization behavior. We
therefore present results for the singular power-law model, which were
calculated using the deflection angles and magnification matrices
derived by Barkana (1998) and implemented in the FASTELL software
package.

Plotted in Fig.~6 are the cumulative distributions of $r_{12}$ and
$r_{13}$ for the Moore profile. The unbiased fractions of Moore
triples with $r_{12} < 1.5$, $2.0$ and $5.0$ are, respectively,
$0.99$, $1.00$ and $1.00$ for $\gamma = 0$; $0.45$, $0.71$ and $0.97$
for $\gamma = 0.05$; $0.44$, $0.69$ and $0.98$ for $\gamma = 0.10$;
and $0.59$, $0.77$ and $0.95$ for $\gamma = 0.15$. The unbiased
fractions of Moore triples with $r_{13} < 20$, $50$ and $100$ are,
respectively, $0.32$, $0.49$ and $0.61$ for $\gamma = 0$; $0.33$,
$0.51$ and $0.64$ for $\gamma = 0.05$; $0.44$, $0.69$ and $0.86$ for
$\gamma = 0.10$, and $0.74$, $0.99$ and $1.00$ for $\gamma =
0.15$. The largest difference between the Moore and NFW predictions
for arcsecond-scale triples is due to the naked cusp configurations.
The tangential caustic is less prominent for a Moore model than it is
for an NFW model in the same shear field. Consequently, while naked
cusps comprise a significant fraction of NFW triples for small shears
($\gamma \la 0.05$), they do not contribute at all to the Moore
statistics until the shear grows large ($\gamma \ga 0.10$). But
despite this effect, dark halos described by the steeper mass profile
of Moore et al.\ still produce lensed triples with small magnification
ratios and relatively bright third images.

\section{Comparison to Current Candidates}

A number of dark lens candidates have been reported in the literature,
and their true nature is still a matter of vigorous debate (e.g.,
Hawkins 1997; Kochanek et al.\ 1999; Mortlock et al.\ 1999; Peng et
al.\ 1999). Table~1 lists nine ambiguous quasar pairs. We have
excluded those systems for which the lensing hypothesis has been
rejected based on gross spectral or photometric differences between
the components.  Technically Q2345+007 (Weedman et al.\ 1982) is no
longer ambiguous, as recent X-ray observations have shown it to be a
binary (Green et al.\ 2002), but we include it for historical
reasons. Below we determine whether the component fluxes of these
quasar pairs are compatible with the dark lens hypothesis.

First, dark lens candidates typically consist of two, not three,
detectable images. Bounds on additional images in these systems are
listed in Table~1. The absence of third images down to these levels
($r_{13,{\rm limit}} \ga 50$) would be very unlikely in a candidate
sample dominated by lenses, particularly if dark halos were best
approximated by an NFW model (Figs.~4 and 5). Second, the observed
flux ratios in many of the ambiguous pairs are significantly larger
than unity. Three of the nine pairs have $r_{12,{\rm obs}} > 10$,
which Figs.~2, 3 and 6 show to be virtually impossible for genuine
dark lenses. However, even the quasar pairs with moderate flux ratios
($3 < r_{12,{\rm obs}} < 5$) would be uncommon in the dark lens
hypothesis. Note that the paucity of small flux ratio pairs cannot be
due to the selection function, as their identification is always
favored (Kochanek 1995b).

Quantitative results are presented in Table~2. For each quasar pair
and model, we calculate the unbiased fraction of triples that have a
primary magnification ratio $r_{12}$ smaller than the observed value,
or a third image brighter than the listed bound.  Fractions with the
third image constraint excluded are also shown. Each of the systems is
an outlier in the predicted dark lens distributions, particularly
those for the NFW profile. Five of the nine pairs have $p (r_{12} <
r_{12,{\rm obs}} \vee r_{13} < r_{13,{\rm limit}}) > 0.95$ in all
eight models studied.  These pairs are strongly incompatible with dark
lensing.  Three other pairs have $0.90 < p (r_{12} < r_{12,{\rm obs}}
\vee r_{13} < r_{13,{\rm limit}}) < 0.95$ in each of the models.
QJ0240--343, with an observed flux ratio of $r_{12} = 2.1$, is
moderately compatible with one of the four Moore models, but no NFW
models.  Magnification bias has little effect on the above results.
In conclusion, while no quasar pair can be absolutely rejected as a
dark lens based on probability arguments alone, the above analysis
demonstrates that most current candidates fail to exhibit the
properties expected for dark lenses. Therefore, these pairs are likely
to be binary quasars, as suggested by Kochanek et al.\ (1999).

The only system that meets the qualitative dark lens criteria is APM
08279+5255 (Ibata et al.\ 1999), which consists of three quasar
components with a maximum separation of $0\farcs38$.  The outer two
components carry nearly equal flux ($r_{12,{\rm obs}} \simeq 1.3$),
and the central component is very bright ($r_{13,{\rm obs}} \simeq
5.7$).  The striking similarity of the quasar colors and the presence
of a third compact component means that APM 08279+5255 is almost
certainly a lens. No lensing galaxy has yet been detected, but this is
not a surprise considering the brightness of the quasars ($m_H \sim
13-14$).  Radio observations have recently detected line emission
offset from the position of the three lensed components (Lewis et al.\
2002), which suggests imaging by a naked cusp. Because third images
are rare for the steep mass profiles of galaxy lenses (Wallington \&
Narayan 1993; Rusin \& Ma 2001), the unique morphology of APM
08279+5255 opens the possibility of strong lensing by a shallow dark
matter halo.  Assuming that the system is a standard triple (source
inside the radial caustic), Mu\~noz, Kochanek \& Keeton (2001) have
demonstrated that a cuspy profile even shallower than NFW is necessary
to reproduce the image data.  Our Monte Carlo tests find that the flux
ratios of APM 08279+5255 are moderately consistent with an NFW model
normalized to a present-day concentration of $c_0 \simeq 7-10$, but
only if the quadrupole is small ($\gamma \la 0.05$). Naked cusps
dominate for larger shears, and such systems tend to have values of
$r_{12}$ and $r_{13}$ much smaller than the observed ratios. In
contrast, while Moore triples commonly form with $1.3 < r_{12} < 2.0$
(Fig.~6), the third image tends to be significantly fainter. Hence,
while APM 08279+5255 exhibits the qualitative properties expected for
a dark lens, none of the standard models provide a particularly good
quantitative match to the system.

\section{Discussion}

This paper has investigated the properties of gravitational lens
systems produced by dark matter halos. While galaxy lenses form with a
wide range of magnification ratios and typically have undetectable
central images, dark lenses are expected to have very small
magnification ratios and prominent third images. These characteristics
result from the shallow inner mass profiles of dark halos, as derived
by N-body simulations. Relative component fluxes in candidate dark
lens systems are powerful benchmarks for evaluating their
viability. The majority of current candidates have flux ratios that
differ significantly from unity, and do not feature any third
image. These sources therefore fail to exhibit the expected properties
of genuine dark lenses. While the above arguments cannot absolutely
reject any source as a dark lens, they offer further evidence that
binary quasars dominate the sample.  The only system that meets the
qualitative dark lens criteria is APM 08279+5255 (Ibata et al.\ 1999).

There are several possible criticisms of the results presented in this
paper. First, the dark matter profiles were assumed to be smooth,
despite the fact that simulations predict copious substructure in CDM
halos (e.g., Moore et al.\ 1999a). Substructure can alter the
magnifications of individual lensed images (Mao \& Schneider 1998), an
effect that may have already been detected (Dalal \& Kochanek
2002). However, Monte Carlo tests indicate that substructure tends to
decrease magnification ratios in doubles (Metcalf \& Madau 2001), at
least for isothermal profiles. If this result can be extrapolated to
the shallow halo profiles, then high flux ratio dark lenses may be
even rarer than we predict. Second, we assumed that dark lenses will
have concentration parameters similar to those of the general halo
population. The simulations of Bullock et al.\ (2001), for example,
predict a range of halo concentrations, and lensing will tend to
select more concentrated mass distributions because of their larger
cross sections. Will this make the gravitational potentials of dark
lenses significantly steeper?  Consider NFW halos with concentration
parameters a factor of three larger than their typical values. There
are not many halos found in this range (Bullock et al.\ 2001). Lenses
with $1'' < \Delta \theta < 10''$ then correspond to normalizations of
$0.25 < \kappa_0 < 0.55$ ($0.3 < z_d < 0.7$). Because Figs.~2 and 4
show that such lenses still have very small primary flux ratios and
bright third images, ``concentration bias'' should not alter our
general conclusions.  Finally, one might suggest that the measured
flux ratios may not be good estimators of the true magnification
ratios if these pairs were real lenses, because of either differential
extinction, microlensing, or a conspiracy between source variability
and time delays. This is unlikely to be a major concern,
however. Similarity between the quasar colors (e.g., Tyson et al.\
1986; Hewett et al.\ 1998), often cited in support of lensing, argues
against the extinction hypothesis. Furthermore, the relatively small
levels of variation observed in the flux ratios -- in some cases over
a baseline of more than a decade -- suggest that time delays and
microlensing are not significantly clouding the measurements.

While we have focussed on triples, five-image lens systems are
expected to play an important role in dark lensing. Small quadrupole
perturbations will result in a large tangential caustic for shallow
profiles, thereby producing many quads. Plotted in Fig.~7 are the
unbiased fractions of five-image lenses formed by the NFW model. More
than half of NFW lenses may be quads, even for very small
shears. These fractions will be further increased by magnification
bias. Using the Moore profile, the unbiased quad fractions are $0.06$,
$0.25$ and $0.47$ for $\gamma = 0.05$, $0.10$ and $0.15$,
respectively. The absence of quads makes the current dark lens
candidate sample even more peculiar.

Dark lenses would be powerful probes of the mass distributions in
primordial dark matter halos -- assuming they can be
found. Unfortunately, the shallow profiles mean that CDM halos are
very inefficient lenses (e.g., Wyithe et al.\ 2001). Cases of strong
lensing are thus expected to be rare, even if empty halos were
abundant. Consequently, the best hope for uncovering dark lenses are
large systematic quasar surveys. None were found in the Cosmic Lens
All-Sky Survey (e.g., Myers et al.\ 1999; 2002), as lens galaxies have
been identified in virtually all of the systems (Jackson et al.\ 1998;
Browne et al.\ 2002), and radio spectral data strongly suggest that
the quasar pair B0827+525 (Koopmans et al.\ 2000) is a physical
binary. The Sloan Digital Sky Survey, however, offers new hope in the
search. The expected properties of dark lenses outlined in this paper
should be a useful tool for critiquing any candidates identified
therein.

\acknowledgements

The author is grateful to Chris Kochanek, Chung-Pei Ma, Nick Sarbu and
Stuart Wyithe for many interesting discussions about dark lensing.
This work was supported by NASA grants NAG5-8831 and NAG5-9265.

\begin{deluxetable}{lccccrrl}
\tablecolumns{7} 
\tablecaption{Historically Ambiguous Quasar Pairs}
\tablehead{ 
\colhead{Source} & \colhead{$N_{\rm comp}$} &
\colhead{Type} & \colhead{$z_s$} & \colhead{$\Delta \theta$} & \colhead{$r_{12,{\rm
obs}}$} & \colhead{$r_{13,{\rm limit}}$} & 
\colhead{Reference}} 
\startdata 
MG0023+171    & 2 & $O^2R^2$ & 0.95 & $4\farcs8$ & $\sim 10^a$   & $\sim 20$   	& Hewitt et al.\ 1987\\ 
QJ0240--343   & 2 & $O^2$    & 1.41 & $6\farcs1$ & $2.1$   & $\sim 100^b$    	& Tinney 1995\\ 
B0827+525     & 2 & $O^2R^2$ & 2.06 & $2\farcs8$ & $2.8^c$ & $\sim 100$ 	& Koopmans et al.\ 2000\\ 
Q1120+0195    & 2 & $O^2$    & 1.47 & $6\farcs5$ & $>60$   & $>100$   		& Meylan \& Djorgovski 1989\\ 
LBQS1429--008 & 2 & $O^2$    & 2.08 & $5\farcs1$ & $17.4$  & $>100$    	& Hewett et al.\ 1989\\ 
Q1634+267     & 2 & $O^2$    & 1.96 & $3\farcs8$ & $3.3^d$   & $\sim 50$ 	& Djorgovski \& Spinrad 1984\\ 
Q2138--431    & 2 & $O^2$    & 1.64 & $4\farcs5$ & $3.0$   & $\sim 70$ 	& Hawkins et al.\ 1997\\ 
LBQS2153--2056& 2 & $O^2$    & 1.85 & $7\farcs8$ & $14.5$  & $>100$   	& Hewett et al.\ 1998\\ 
Q2345+007     & 2 & $O^2$    & 2.15 & $7\farcs3$ & $4.0$   &  $>100^e$    & Weedman et al.\ 1982\\ 
\enddata 

\tablecomments{Listed for each pair are the number of quasar
components ($N_{\rm comp}$), radio-brightness classification, source
redshift ($z_s$), angular separation ($\Delta \theta$), observed flux
ratio ($r_{12,{\rm obs}}$), bound on additional components
($r_{13,{\rm limit}}$), and reference. Many of the parameters are
taken from the tables of Peng et al.\ (1999) or Mortlock et al.\
(1999). Radio flux ratios are less susceptible to microlensing and
extinction, and are used where available. Values of $r_{13,{\rm
limit}}$ are from the listed references, except where noted. $^a$ 5
GHz measurement (optical ratio is $\sim 3$); $^b$ CfA-Arizona Space
Telescope Lens Survey; $^c$ 8.4 GHz measurement; $^d$ updated value
from Peng et al.\ (1999); $^e$ Tyson et al.\ (1986).}
\end{deluxetable}

\clearpage

\begin{deluxetable}{lccc|ccc}
\tablecolumns{7}
\tablecaption{Quasar Pairs vs. the Dark Lens Distribution}
\tablehead{ 
\colhead{}    &  \multicolumn{3}{c}{NFW profile} &   \multicolumn{3}{c}{Moore profile} \\ 
\colhead{Source} & \colhead{$\gamma = 0.05$} & \colhead{$\gamma =
0.10$}& \colhead{$\gamma = 0.15$} & \colhead{$\gamma = 0.05$} & \colhead{$\gamma = 0.10$} &
\colhead{$\gamma = 0.15$}}
\startdata 
MG0023+171      &0.99 (0.99)  &1.00 (0.99) & 1.00 (1.00) &1.00 (1.00) &1.00 (1.00) &0.99 (0.99)\\
QJ0240--343     &1.00 (0.63)  &1.00 (0.65) & 1.00 (1.00) &0.85 (0.74) &0.96 (0.72) &1.00 (0.78)\\
B0827+525       &1.00 (0.76)  &1.00 (0.92) & 1.00 (1.00) &0.91 (0.85) &0.97 (0.86) &1.00 (0.86)\\
Q1120+0195      &1.00 (1.00)  &1.00 (1.00) & 1.00 (1.00) &1.00 (1.00) &1.00 (1.00) &1.00 (1.00)\\
LBQS1429--008   &1.00 (1.00)  &1.00 (1.00) & 1.00 (1.00) &1.00 (1.00) &1.00 (1.00) &1.00 (1.00)\\
Q1634+267       &1.00 (0.83)  &1.00 (0.92) & 1.00 (1.00) &0.92 (0.90) &0.96 (0.93) &0.99 (0.90)\\
Q2138--431      &1.00 (0.80)  &1.00 (0.87) & 1.00 (1.00) &0.91 (0.87) &0.96 (0.90) &0.99 (0.88)\\
LBQS2153--2056  &1.00 (1.00)  &1.00 (0.99) & 1.00 (1.00) &1.00 (1.00) &1.00 (1.00) &1.00 (1.00)\\
Q2345+007       &1.00 (0.90)  &1.00 (0.91) & 1.00 (1.00) &0.95 (0.94) &0.98 (0.96) &1.00 (0.93)\\
\enddata

\tablecomments{Listed for each quasar pair and model are the predicted
fractions of triples $p (r_{12} < r_{12,{\rm obs}} \vee r_{13} <
r_{13,{\rm limit}})$ and, in parentheses, $p (r_{12} < r_{12,{\rm
obs}})$. These probabilities are $1.00$ ($1.00$) for all pairs in
spherical ($\gamma = 0$) models. For the NFW profile, the fractions
were calculated by deriving $\kappa_0$ from the angular splitting
scale $\Delta \theta$, assuming a lens redshift $z_d = 0.5$, a source
redshift $z_s = 2.0$ (measured redshifts were ignored for simplicity),
a present-day concentration parameter $c_0 = 7$, and a flat
$\Omega_{\Lambda} = 0.7$ cosmology. Varying these parameters within
their reasonable ranges has a negligible effect on the results. All
the model predictions are unbiased. Including bias has little
qualitative impact on the derived fractions.}

\end{deluxetable}

\clearpage

\clearpage
\begin{figure*}
\begin{tabular}{cc}
\psfig{file=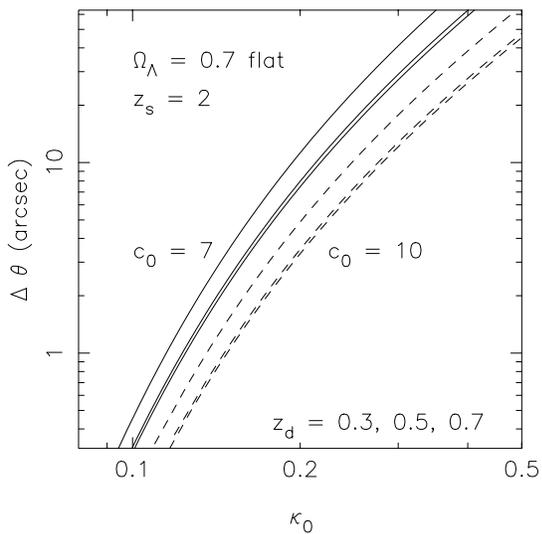}&
\end{tabular}
\figurenum{1}
\caption{Conversion between the NFW normalization ($\kappa_0$) and
angular splitting scale ($\Delta \theta$) for fixed source redshift
$z_s = 2$, present-day concentration parameters $c_0 = 7$ (solid) and
$10$ (dash), and lens redshifts $z_d = 0.3$, $0.5$ and $0.7$ (left to
right). A flat $\Omega_{\Lambda} = 0.7$ cosmology is assumed. The
separation $\Delta \theta$ is approximated as twice the size of the
tangential critical curve for the spherical case. Naked cusp
configurations tend to have somewhat smaller image separations than
standard configurations produced by the same deflector. However, the
small correction does not affect the general trends discussed in this
paper, and is therefore ignored. Note that arcsecond-scale lenses are
dominated by $0.1 < \kappa_0 < 0.2$.}
\end{figure*}

\clearpage
\begin{figure*}
\begin{tabular}{c}
\psfig{file=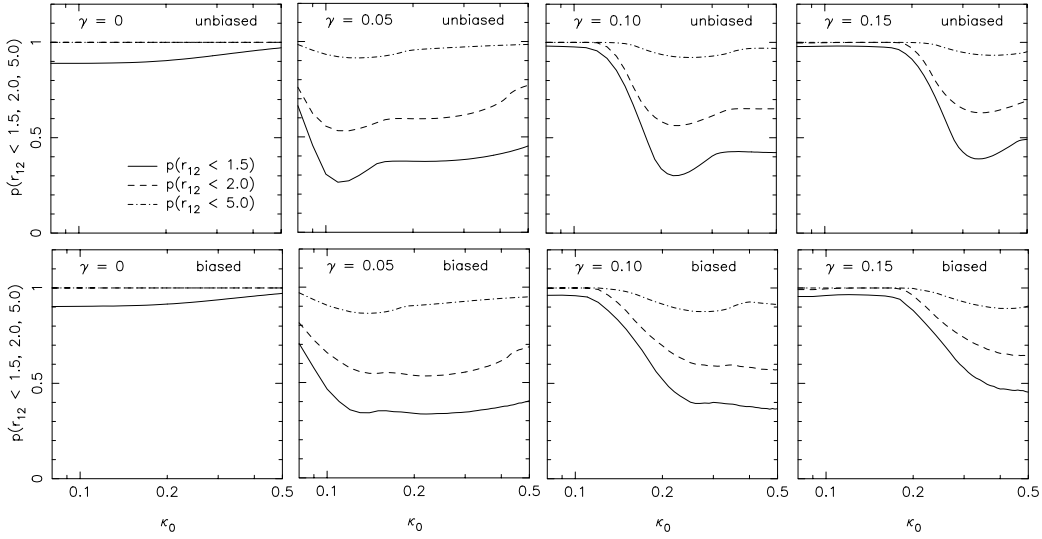}\\
\end{tabular}
\figurenum{2}
\caption{Primary magnification ratio $r_{12} \equiv |\mu_1/\mu_2|$ in
NFW triples: fractions ($p$) of triples with $r_{12} < 1.5$ (solid),
$2.0$ (dash) and $5.0$ (dot-dash) as a function of $\kappa_0$ for
shear amplitudes $\gamma = 0$, $0.05$, $0.10$ and $0.15$. Unbiased and
biased values are shown.}
\end{figure*}

\begin{figure*}
\begin{tabular}{c}
\psfig{file=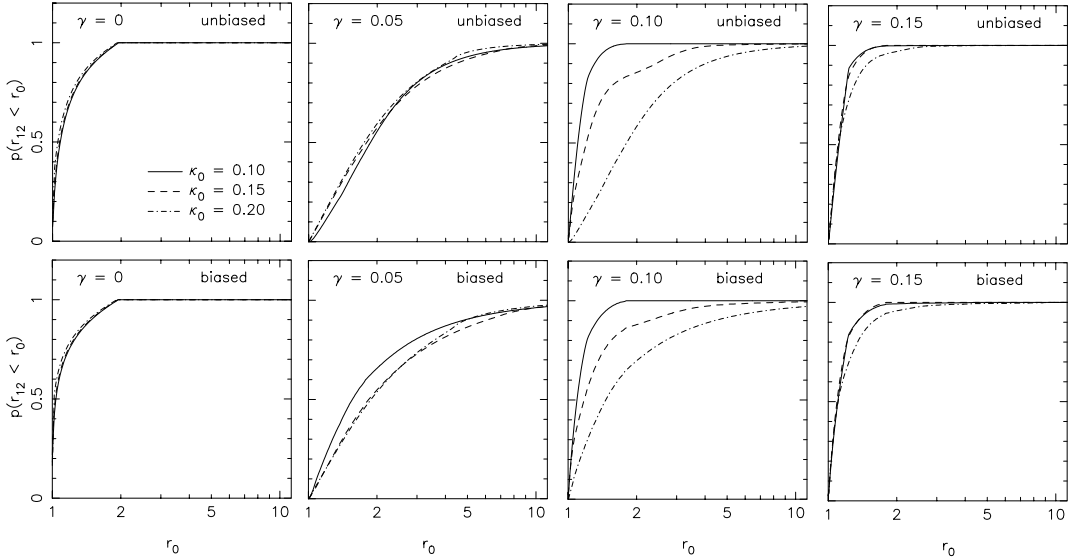}\\
\end{tabular}
\figurenum{3}
\caption{Distribution of magnification ratios $r_{12}$ in NFW triples:
fractions $p(r_{12} < r_0)$ for $\kappa_0 = 0.10$ (solid), $\kappa_0 =
0.15$ (dash) and $\kappa_0 = 0.20$ (dot-dash). Unbiased and biased
values are shown for shear amplitudes $\gamma = 0$, $0.05$, $0.10$ and
$0.15$.}
\end{figure*}

\clearpage
\begin{figure*}
\begin{tabular}{c}
\psfig{file=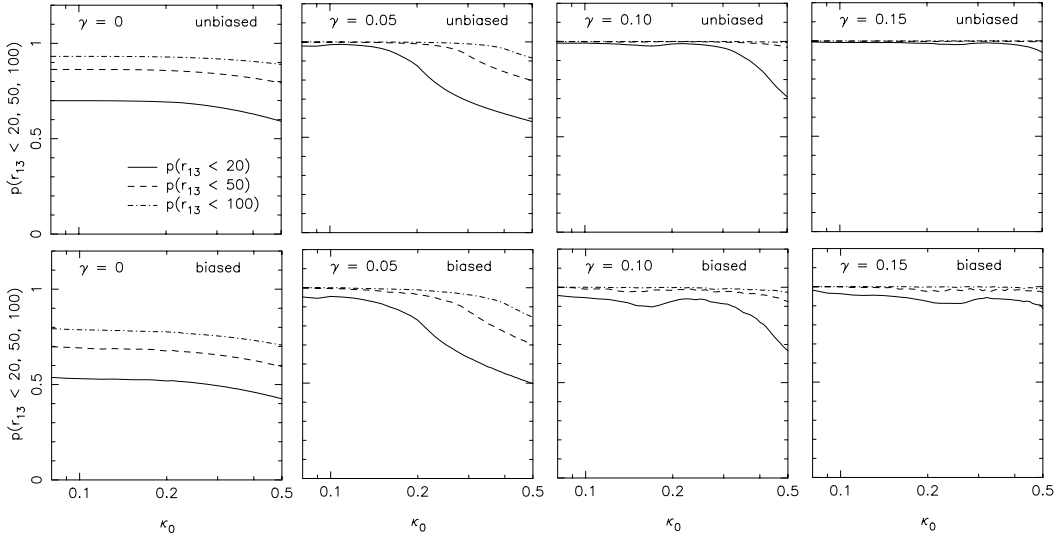}\\
\end{tabular}
\figurenum{4}
\caption{Same as in Fig.~2, but for the brightness of the third image
in NFW triples ($r_{13} \equiv |\mu_1/\mu_3|$): fractions ($p$) of
triples with $r_{13} < 20$ (solid), $50$ (dash) and $100$ (dot-dash).}
\end{figure*}

\begin{figure*}
\begin{tabular}{c}
\psfig{file=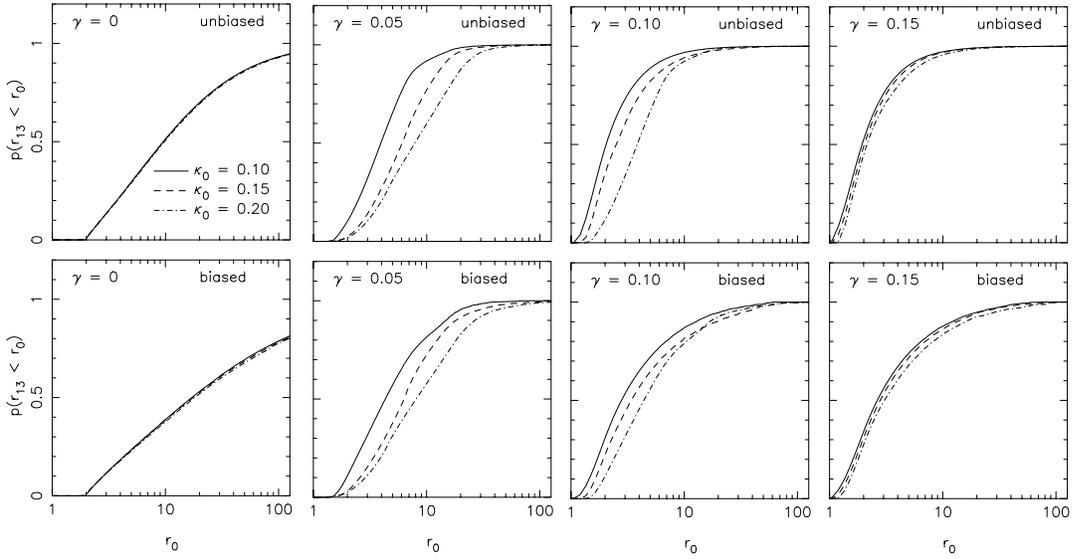}\\
\end{tabular}
\figurenum{5}
\caption{Same as in Fig.~3, but for ratio $r_{13}$: fractions
$p(r_{13} < r_0)$.}
\end{figure*}

\clearpage
\begin{figure*}
\begin{tabular}{c}
\psfig{file=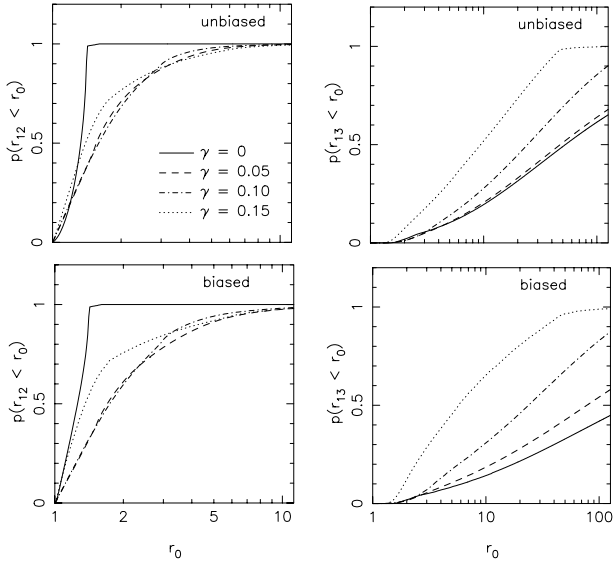}\\
\end{tabular}
\figurenum{6}
\caption{Distribution of magnification ratios $r_{12}$ (left) and
$r_{13}$ (right) in Moore triples. Plotted are the fractions
$p(r_{12,13} < r_0)$ for $\gamma = 0$ (solid), $0.05$ (dash), $0.10$
(dot-dash) and $0.15$ (dot).  Unbiased and biased values are shown.}
\end{figure*}

\begin{figure*}
\begin{tabular}{c}
\psfig{file=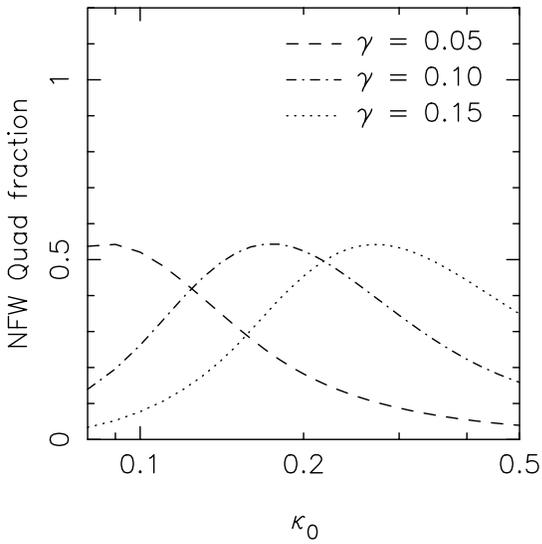}\\
\end{tabular}
\figurenum{7}
\caption{Fraction of five-image systems produced by the NFW model as a
function of $\kappa_0$ for $\gamma = 0.05$ (dash), $0.10$ (dot-dash)
and $0.15$ (dot). Only unbiased values are shown. Note that the curves
turn over for small $\kappa_0$. In this regime the number of quads is
diluted, as some sources within the tangential caustic are now lensed
into cusp configurations. Consequently, deflectors with $\kappa_0$ to
the left of the above peaks produce triples that are dominated by
naked cusps.}
\end{figure*}

\end{document}